# Towards New Requirements Engineering Competencies

Sami Jantunen[1], Rex Dumdum[2], Donald C. Gause[3]
[1] School of Engineering Sciences, LUT University, Lappeenranta, Finland
[2] School of Business and Global Innovation, Marywood University, Scranton, PA, USA
[3] Binghamton University and Savile Row LLC, USA
sami.jantunen@lut.fi, rexdum2@marywood.edu, doncgause@gmail.com

*Abstract*— **Many of the requirements engineering (RE) difficulties have been argued to be due to the evolving nature of design problems in dynamic environments, characterized by high levels of uncertainty, ambiguity and emergence. It has also been argued that these challenges cannot be solved by focusing primarily on notations, tools, and methods. The purpose of this vision paper is to understand better what kinds of new competencies are needed when expanding RE practices to cope with complex systems in dynamic environments. We intend to achieve our goal by discussing: 1) how increased complexity affects RE practices, and 2) what viewpoints have been found most salient when aligning RE practices with the design problem at hand. Based on our findings, we argue for the importance of contextual intelligence, the ability to recognize and diagnose contextual factors and then intentionally and intuitively adjust behavior. We also outline some of the important competencies that need to be developed for future RE practitioners to deal with complex problems.**

*Keywords- Requirements Engineering, Complexity*

## I. INTRODUCTION

The environment in which Requirements Engineering (RE) is practiced has changed dramatically throughout the years of software engineering [1]. In the early days (1960s and 1970s), design problems had strong links with science and mathematics, seeding the belief that there is one reality that is measurable and essentially the same for all. There were originally no generic methods or techniques for bridging the gap between user needs and technology capabilities [2, p. 4]. This was the era of *just building the software* [3]. Since then, RE has evolved towards *designing* (1970s and 1980s), and *defining the functionality* (1990s and 2000s) [3]. Consequently, the task of software engineers has largely become to design and implement software that models the single reality that all stakeholders share. Such assumptions have laid the foundation for a professional paradigm that is deeply entrenched and still dominant today. Software engineers have learned to see themselves as being *experts* in technologies, tools, development methods and project management [4]. This has resulted in the desire to design software in a systematic, formal and rational manner.

Now that the focus in RE has started to shift towards *understanding and defining the context* [3], organizations are increasingly recognizing the need for a broader domain of inquiry including the conflicts and complexities of design problems. Over the years, the idea of requirements has changed from "single, static and fixed-point statements of desirable system properties into dynamic and evolving rationales that mediate dynamic change between the business environments and the design and implementation worlds" [5]. Yet, RE research has persistently developed a variety of formal and computational models for reasoning about requirements, while remaining largely atheoretical in its view of RE as a socio-technical endeavour [6].

Much of the RE-related difficulties have been argued to result from the struggles of the research community to keep pace with the evolving nature of design problems [6, 7]. In order to keep up with the changing world, we need to take a fresh perspective to RE where "we intertwine requirements and contexts, evolve designs and ecologies, manage through architectures and learn to recognize and mitigate against design complexity" [5]. Jarke and his colleagues [5] have argued that "answers to this challenge cannot come just from doing more of the same—i.e., traditional RE research focusing primarily on notations, tools, and methods". "The RE field needs to also carefully evaluate some of its sacred assumptions and as a result its research scope may have to become more interdisciplinary" [5].

The purpose of this vision paper is to understand better what new competencies RE practitioners need for developing complex systems in dynamic environments. To this end, we discuss: 1) how increased complexity affects RE practices (Section II), and 2) what viewpoints have been found most salient when aligning RE practices with the design problem at hand (Section III). In Section IV, we reflect our findings and argue for the importance of contextual intelligence. Finally, Section V concludes the paper by outlining some of the important competencies that need to be developed for future RE practitioners to deal with complex problems.

## II. HOW COMPLEXITY AFFECTS RE

Much of the traditional scientific work has been built on the supposition that the unknowability of situations is the result of a lack of information [8]. This has led to an emphasis on uncertainty reduction through ever-increasing information seeking and processing, with a central goal of avoidance of surprise [8]. When dealing with complex systems, however, surprises are inevitable. According to the complexity view, products, whose requirements need to be managed, co-evolve with their environment, emerging to satisficing solutions, causing surprises and changes along the way. Whereas most of the RE interest in the past has focused on understanding and managing the *inner* and *static* complexity of the design task by using abstraction, modularization, and related principles, today's complexity is also *external* and *dynamic* [1]. As a result, RE today is significantly different in terms of the requirements volatility, vagueness, variance, and velocity [1]. This creates a "need to expand and deepen the range of theoretical frameworks that help conceptualize new RE complexity and generate strategies to mitigate its effects" [1].

### III. DETERMINING AN ADEQUATE FIT BETWEEN RE PRACTICES AND A DESIGN PROBLEM

The objective of determining adequate fit between RE tasks and the design problem at hand has been argued to be a matter of considering three primary aspects: 1) the nature of a design problem, 2) the nature of the environment, and 3) the beliefs of the requirements engineer [9]. We explore these aspects in greater detail below.

*A. Identifying the nature of the design problem*

The reason for identifying the nature of the design problem is to understand better which RE approaches are likely to succeed in dealing with the problem at hand [9]. One potential approach is to make a distinction between *simple*, *complicated*, *complex*, and *chaotic* problems, as suggested by the Cynefin framework [10]. According to the framework, *simple* and *complicated* problems assume an ordered universe, where cause-and-effect relationships are perceptible, and right answers can be determined based on the facts [10]. These are the problem domains for which traditional RE approaches have proved to be particularly effective. *Complex* and *chaotic* problems are unordered— there is no immediately apparent relationship between cause and effect, and the way forward is determined based on emerging patterns [10].

Unordered problems require different, often counter-intuitive, responses [10]. Requirements can't be ferreted out when dealing with *complex* problems. This is the domain to which much of contemporary business has shifted [10]. We can understand why things happen only in retrospect and thus an iterative and experimental approach is needed. Instructive patterns can emerge if the practitioners conduct experiments that are safe to fail. RE practitioners need to probe first, then sense, and then respond [10]. The attempt to over-control RE activities would preempt the opportunity for informative patterns to emerge [10]. When dealing with *chaotic* problems, searching for right answers would be pointless. The relationships between cause and effect shift constantly. No manageable patterns exist—only turbulence. In this domain, practitioners' immediate job is to first establish order, then sense where stability is present and from where it is absent, and then respond by working to transform the situation from chaos to complexity, where the identification of emerging patterns can both help prevent future crises and discern new opportunities [10].

*B. Assessing the nature of the environment*

This assessment investigates whether the environment for the system is stable or dynamic. The challenges caused by dynamic environment are often related to *requirements volatility*, the growth or changes in requirements during a project's development lifecycle [11]. Traditional RE approaches have largely based on a belief that requirements need to be spelled out at the beginning of a project and that changes should be avoided, if possible [11, 12]. Reifer [12] have argued that "The only thing wrong with these techniques is that they don't work in today's environments". It is now widely understood that requirements change is unavoidable and that requirements don't necessarily change because of poor RE practices [13]. We need better awareness to the strengths and weaknesses of different process models for handling volatile requirements. In volatile environments RE need to be seen more as a "*learning*, rather than as a gathering, process" [12], where changes can be expected and are welcomed throughout the development [14].

*C. Understanding the beliefs of the requirements engineer*

People can have diverging perspectives of what kinds of RE methods, tools, techniques and approaches are considered "appropriate". Understanding the differences in perspectives is critical, since these differences often lies at a deeper philosophical underpinnings [9]. Much of the differences seem to be related to the question of whether RE is fundamentally an objective or subjective endeavour. From the objective perspective, nailing down requirements early and with concreteness is a laudable goal [9]. This leads RE to focus on the creation of static, well defined understanding (not subject to differential interpretations), emphasizing on the importance of basing RE approaches upon a systematic protocol and technique [9]. However, practitioners taking subjective perspective are likely to think that requirements are "based on shifting knowledge and circumstances, variability in interpretations, with a goal of refinement of requirements perspectives" [9]. This view calls for a requirements engineer to be creative and less deterministic since complex situations are fluid and uncertain [9].

The intention of examining one's philosophical underpinnings is to ensure that the practitioners' beliefs do not contradict with the nature of the design problem. Inability to recognize the differences in beliefs is likely to create conflict in RE for complex situations [9]. Unfortunately, philosophical incongruities often come about through other conflicts (e.g., attacks claiming poor RE due to continual shifting interpretations inherent in complex situations) [9]. Finding consensus in such situations is difficult because it would mean asking someone to admit inadequacy of one's mental map and to surrender and make adaptations on one's deeply held belief system [9].

### IV. TOWARDS CONTEXTUAL INTELLIGENCE

Most RE practices play out differently in different contexts [15]. Hence, determining effective ways to deal with a given design problem requires *contextual intelligence*: the ability to recognize and diagnose the plethora of contextual factors inherent in an event or circumstance and then intentionally and intuitively adjust behavior in order to exert influence in that context [16]. Effectiveness can be achieved when compatibility is established between RE approaches and context [9]. We have in Section III discussed three aspects that are salient in increasing the chances for RE to effectively deal with a given design problem. First, the nature of the design problem should be taken into account. If the problem is simple or complicated [10], it should be treated with traditional RE processes, as those are proven and are appropriate for this type of problems [9]. If, however, the design problem is complex or chaotic [10], practitioners should be wary of the use of traditional RE approaches as these have not been designed for and do not fully appreciate complex situations [9]. These unordered problems require problem-solving approaches that are more experimental by nature [10]. Second, practitioners should understand that, in dynamic situations, they should maintain flexibility in RE tasks and ensure that requirements can, and will, change as new knowledge and understanding of the system and context emerge [9]. Assuming that RE will yield stable, verifiable, definitive, and objective requirements for dynamic situations need to be challenged [9]. Finally, the compatibility of requirements engineers' beliefs need to be considered. When dealing with complex situations, one needs to accept: 1) emergence, ambiguity and uncertainty; 2) the transient nature of requirements based on shifting knowledge and circumstances, variability in interpretations; and 3) a need to be creative and less deterministic.

Some within RE community have already recognized the importance of contextual intelligence. In their 2000 RE Roadmap Paper, Nuseibeh and Easterbrook [17] identified the emergence of three radical new ideas that challenged and overturned the orthodox views of RE: 1) the idea that modelling and analysis cannot be performed adequately in isolation from the organizational and social context in which any new system will have to operate; 2) the notion that RE should not focus on specifying the functionality of a new system, but instead should concentrate on modelling indicative and optative properties of the environment; and 3) the idea that the attempt to build consistent and complete requirements models is futile, and that RE has to take seriously the need to analyze and resolve conflicting requirements, to support stakeholder negotiation, and to reason with models that contain inconsistencies. In a similar vein, Gause [18] has argued that our understanding of context defines our view of the design problem. He further argues that anything that we can do to make our design thinking and processes more visible will give us an improved look at relevant context issues, resulting in more complete design and better fit of form to context.

In order to improve contextual intelligence, we not only need ways to understand the nature of contexts and to know when traditional approaches are beyond the limits of their applicability, we also need to develop new RE approaches for dealing with complex situations [9]. This may, however, turn out to be a difficult task. Past success with existing approaches often creates incredible obstacles to adapting or responding to changing contexts [19]. People can be strongly biased by their existing knowledge and rarely can interpret what they see without that bias [19]. This phenomenon is known as "*Paradigm Paralysis*" [20, p. 65]. We believe that contextual intelligence can help to overcome paradigm paralysis. The first steps in that adaptation are the toughest: jettisoning assumptions about what will work and then experimenting to find out what actually does work [15].

Progress towards dealing with complex problems is already underway. As an example, *agile RE* has already acknowledged that rapidly changing business environment challenges traditional RE practices [14]. Agile practices, however, might not be the final step of software development. Olsson et al. [21] argue that software development companies are moving beyond the concept of agile development towards *R&D as innovation experiment system*, where "requirements evolve in real-time based on data collected from real-time customer use instead of being frozen early based on the opinions of product management about future customer needs". To our understanding, this kind of software development approach takes a step towards experimentation, learning, empowerment of practitioners and emergence of solutions, all of which have been suggested strategies to deal with complexity [22, 23].

V. TOWARDS NEW COMPETENCIES

We have in this paper argued for the importance of contextual intelligence, the ability to understand the nature of context and the corresponding RE approaches that are likely to be effective in that particular context. We have also argued that RE is increasingly facing complex design problems, for which there is a need to develop new kinds of approaches. Studies of complexity (e.g. [8, 22, 24]) consistently argue that complexity is best tackled with strategies such as *experimentation*, *learning*, *self-organization, emergence* and *empowered action*. These strategies significantly change the role of a requirements engineer from one obeying and following processes towards one that is attentive to cues and empowered to react.

When Jarke and his colleagues [5] identified principles to guide future RE research, they also outlined an extensive list of research topics and questions, addressing challenges that are to be answered through new kinds of methods and processes. We argue that this alone will not be enough in complex situations that are to be dealt by empowered stakeholders. When developing competencies for future RE practitioners, we not only need to pay attention to *horizontal development* (the development of technical skills that can be transmitted), but also to *vertical development* (personal skills that must be earned for oneself) [25]. Much of vertical development is about helping individuals to think in more complex ways [25]. To this end, it would be worthwhile to explore disciplines such as organizational studies, complexity sciences, leadership studies, psychology and educational studies. Some of the important competencies increasing contextual intelligence and ability to act in complex situations include:

1. *Learning to learn* [26]. We tend to have very persistent mental models that are not rooted in the facts and that hinder progress [15].
2. *Sensemaking* [27], the ability of: 1) coming up with a plausible understanding; 2) testing this understanding with others through data collection, action, and conversation; and then 3) refining or abandoning the understanding depending on how credible it is [28].
3. *Dialogue*, the skill of using the energy of peoples' differences to enhance collective wisdom by: 1) speaking their true voice, and encouraging others to do the same;

2) listening as a participant; 3) respecting the coherence of others' views; and 4) suspending their certainties [29].
4. *Mindfulness*, the commitment to be attentive and preoccupied with failures in the system and being reluctant to simplify the interpretations of failures [30].
5. *Facilitative leadership* [31], being able to: 1) use active listening skills including paraphrasing, summarizing, reflecting, and questioning; 2) encourage and generate participative discussion in groups; 3) help stimulate creative thinking through idea-generation processes; 4) stimulate strategic consideration of alternatives and informed decision-making of appropriate choices, 5) manage contrasting perspectives and opinions that might result in conflict among members of a group, 6) help individuals and groups reflect on their experiences and capture relevant learning, 7) help shape more powerful and strategic questions for exploration.

The need for vertical competence development is particularly discussed in the leadership literature. It appears to us that the disciplines of leadership and RE are coming closer. While leadership research is moving towards models, where leadership is distributed more evenly across the organization [32] and where leadership is more adaptive and facilitative by nature [31], future requirements engineers, dealing with complexity, need to be attentive, empowered, and able to work collaboratively. In other words, leadership is increasingly required from the RE practitioners when problems are novel or have not been experienced before [33].